# Defect recombination induced by density-activated carrier diffusion in nonpolar InGaN quantum wells

Fan Yang,[1] Chunfeng Zhang,[1,a] Chentian Shi,[1] Min Joo Park,[2] Joon Seop Kwak,[2,b] Sukkoo Jung,[3] Yoon-Ho Choi,[3] Xuewei Wu,[1] Xiaoyong Wang,[1] and Min Xiao[1,4,c]

[1]National Laboratory of Solid State Microstructures and Department of Physics, Nanjing University, Nanjing 210093, China
[2]Department of Printed Electronics Engineering, Sunchon National University, Sunchon, Jeonnam 540-742, Korea
[3]Emerging Technology Laboratory, LG Electronics Advanced Research Institute, Seoul 137-724, Korea
[4] Department of Physics, University of Arkansas, Fayetteville, Arkansas 72701, USA

**Abstract:**

We report on the observation of carrier-diffusion-induced defect emission at high excitation density in a-plane InGaN single quantum wells. When increasing excitation density in a relatively high regime, we observed the emergence of defect-related emission together with a significant reduction in bandedge emission efficiency. The experimental results can be well explained with the density-activated carrier diffusion from localized states to defect states. Such a scenario of density-activated defect recombination, as confirmed by the dependences of photoluminescence on the excitation photon energy and temperature, is a plausible origin of efficiency droop in a-plane InGaN quantum-well light-emitting diodes.

[a] cfzhang@nju.edu.cn
[b] jskwak@sunchon.ac.kr
[c] mxiao@uark.edu

The efficiency droop phenomenon, i.e., the drop in quantum efficiency of light emission at high current injection, is a major obstacle for the high-brightness, high-efficiency applications of InGaN blue-green quantum-well (QW) light-emitting diodes (LEDs).[1-3] In the past decade, multiple mechanisms have been proposed to explain the efficiency droop including Auger recombination induced by multi-particle interactions,[4-8] electron leakage due to poor hole injection and electron overflow,[9,10] carrier loss related to reduction or saturation of carrier localization,[11,12] and other processes including the density-activated defect recombination (DADR).[13,14] Some of these factors have been found to be tightly associated with the internal polarization field.[15-21] The reduction on internal polarization and the increase of hole concentrations in nonpolar QWs can efficiently suppress the loss during carrier injection.[19-21] Potentially, the interplay of polarization fields and Auger recombination in nonpolar QWs can be manipulated to optimize the device performance.[16] These advantages make nonpolar QWs particularly suitable to achieve low droop devices. Recently, vital progresses have been achieved.[19-21] Upon raising the current density to a level of 100 A/cm$^2$, efficiency droop has been significantly reduced from 50% with polar QWs to 13% in nonpolar QWs.[21] Understanding the mechanism of the residual droop in nonpolar LEDs will be instrumental towards droop-free LED devices.

Besides the intensively-studied Auger recombination, defect-related Shockley-Read-Hall (SRH) recombination is another major non-radiative process.[3] Although it limits the maximum efficiency of LED devices, the SRH process has been excluded from the major mechanisms of the efficiency droop in the early studies since similar droops were observed in samples with very different defect densities.[4,22] Nevertheless, some recent works have proposed high carrier density may activate the defect-related recombination.[13,14] This process of DADR is predicted to be responsible for the efficiency droop.[18,19]

In this letter, we report experimental evidences that the process of DADR can be a major channel responsible for the efficiency droop in nonpolar a-plane InGaN QWs.

We observed the emergence of defect-related emission ($I_{Def}$) together with a significant reduction in blue emission efficiency at bandedge ($I_{Blue}$) when increasing excitation density in a relatively high regime. Besides the electron-electron (e-e) scattering,[14] we argue that another process of thermally-activated carrier diffusion can also drive migration of electrons from localized states to defect states, which is supported by the dependences of emission intensity ratio ($I_{Def}/I_{Blue}$) on the excitation photon energy and temperature. These results suggest the process of DADR as a major factor for efficiency droop at high carrier density, which, in principle, can be avoided by manipulating the lateral carrier diffusion in a-plane InGaN LEDs.

The nonpolar a-plane QW samples were grown on r-plane sapphire substrates consisting of 4 μm thick GaN buffer layer, 2 μm thick n-GaN layer, 15 nm thick InGaN single QW layer, and 150 nm thick p-GaN capping layer. For PL measurements, we employed a femtosecond Ti:sapphire regenerative amplifier (pulse duration ~ 90 fs, Libra, Coherent) and a cw diode laser at 375 nm as excitation sources. For wavelength-dependent measurements, we detuned the excitation wavelength with an optical parametric amplifier (Opera Sola, Coherent). The emission spectra were collected by a spectrograph (Sp 2500i, Princeton Instruments) equipped with a charge-coupled device cooled by liquid nitrogen. The excitation residual was eliminated by an ultrasteep long-pass filter (BLP01-405R-25, Semrock). Temperature-dependent measurements were carried out by using an optical cryostat (MicroCryostatHe, Oxford). The carrier density was changed by detuning excitation pulse energy. For fluence-dependent and temperature-dependent experiments, the wavelength of pulse excitation was set to be 400 nm unless otherwise specified.

In InGaN LEDs, the lateral carrier diffusion is strongly confined owing to the excitation localization effect induced by the fluctuations of well-thickness and well-composition.[23,24] Consequently, the emission properties in InGaN LEDs are quite insensitive to the defects under weak excitation[25] in consideration of the high concentrations of point defects and threading dislocations.[26] Some SRH centers are surrounded by potential barriers, minimizing the possibility of defect recombination in the low density regime. Here, we propose that the thermally-activated diffusion process can overcome the potential barriers and assist the carrier recombination through defect states in high carrier density regime as depicted in Fig. 1(a). Previously, thermal effect was observed to induce carrier diffusion between the weakly and strongly localized states (process A in Fig. 1(a)) in InGaN QWs.[27,28] This process possesses the possibility to populate the defect states (process B in Fig. 1(a)). With increasing carrier density, the state filling effect will reduce the energy barrier required to be overcome, leading to a pronounced defect-related recombination. The possibility of defect recombination in this scenario will be superlinearly dependent on carrier density, leading to the efficiency droop in LED devices.

When radiative processes take part in the defect-related recombination, PL emission related to defect states can be monitored to evaluate the process of DADR. As displayed in Fig.1(b), together with the blue bandedge emission, a broadband yellow emission can be detected from the a-plane QW samples under relatively intense pulse excitation. Such a yellow emission band is known to come from deep-level recombination introduced by defects.[29,30] While the exact mechanisms of deep-level defect states are still under debate[31] and beyond the scope of this work, the dependence of deep level emission on carrier density, as we analyzed below, can be used to study the efficiency droop.

PL spectra under different fluence excitations are shown in Fig. 2(a). It is found that at very low excitation density, i.e., weak cw excitation, defect emission is undetectable (inset, Fig.2(a)). When excitation fluence increases, the defect emission emerges and becomes more important with increasing excitation density. The PL spectra can be roughly fitted by two emission bands with Gaussian profiles (Fig. 1(b)). The fluence-dependent integrated intensities of the two bands are compared in Fig. 2(b). Below the excitation fluence of ~ 50 μJ/cm$^2$, the blue emission intensity is proportional to the excitation fluence. The slope of fluence-dependence drops at higher density, indicating a decrease in light emission efficiency. Simultaneously, the slope for fluence-dependence of defect emission shows a slight increase. This difference is highlighted by the intensity ratio of $I_{Def}/I_{Blue}$. As shown in Fig. 2(c), this ratio shows a critical behavior for fluence dependence which dramatically increases with excitation beyond ~ 50 μJ/cm². These results indicate the defect recombination plays a more important role for carrier loss when carrier density increases. For PL characterization, the normalized bandedge emission intensity per unit excitation power is often used as a metric to evaluate the efficiency droop.[6] This metric is a direct reflection of internal quantum efficiency in the excitation density regime where saturation absorption effect is negligible. As shown in Fig. 2(c), with excitation fluence beyond a similar threshold, a dramatic decrease of the quantum efficiency appears (i.e., the emergence of the efficiency droop). The critical density-dependence occurs at a similar density threshold for both the intensity ratio ($I_{Def}/I_{Blue}$) and quantum efficiency, suggesting the DADR as a plausible factor that induces the efficiency droop.

The fluence dependence of the defect emission observed here is in consistence with the scheme proposed in Fig. 1(a). The superlinear dependence can be naturally explained by the effect of state filling.[23,32] In principle, the state filling would induce a blue shift of the PL emission with increasing excitation fluence,[32] which is clearly seen in the emission spectra (Fig. 2(a)). With increasing carrier density, the possibility for electrons occupied at the weakly-localized states (higher energy levels) increases. The reduced potential barriers can be more easily overcome by the thermally-activated carrier diffusion. Besides the thermally-activated carrier diffusion, several other processes may also make contributions. (1) The process of phonon-assisted hole transport contributes via tunneling at defect sites.[33] This process can be excluded here since it is insensitive to the carrier density. (2) The Auger process induces hot carriers that may relax to the defect states.[8] This process is very sensitive to the carrier density but insensitive to the degree of localization. (3) The e-e scattering can drive the carriers cross the barriers as proposed by Hader *et al*.[13,14] This process is temperature insensitive; however, a density threshold is required to ensure the scattering magnitude sufficient to overcome the potential barriers.[14] To confirm the process of DADR and check the major mechanism responsible for defect recombination, we study the PL properties by changing the excitation photon energy and the temperature.

Figure 3(a) shows the emission spectra recorded upon excitation at different wavelengths with same fluence. Defect emission becomes more important with shorter-wavelength excitation. Such a significant change cannot be explained by the slight increase in light absorption at the shorter wavelength since the blue bandedge emission becomes less dominant. For a quantitative analysis, we plot the value of $I_{Def}/I_{Blue}$ as a function of the excitation photon energy, which behaves differently in two regimes. It slowly increases with increasing photon energy to ~ 3.27 eV, and then rapidly increases with higher energy of excitation photons. These results indicate the degree of localization is an important factor for the defect recombination. At higher states, the possibility of the electron migration to the defect states is higher; when the energy is higher than the potential barriers, defect recombination becomes dominant. These behaviors are explainable within the scheme of DADR, induced by either the e-e scattering and the thermally-activated carrier diffusion, but in contrast to the hot carrier relaxation related to Auger processes. The strong dependence of $I_{Def}/I_{Blue}$ on the excitation photon energy also suggests a high electron kinetic energy at the conduction band favors the defect recombination. In comparison with optical excitation, the electron kinetic energy in an electrically-driven device is higher with increasing current density by applying higher voltage.[8] Thus, the DADR would play a more important role in electrically-driven LEDs.

We further examined the role of thermal effect by temperature-dependent measurements. Figure 4 plots the normalized PL spectra acquired at different temperature. The ratio $I_{Def}/I_{Blue}$ increases with temperature, which could be caused by the thermally-activated carrier diffusion. However, it is a necessary but not a sufficient evidence for the presence of carrier diffusion, since light emission efficiency for both the localized and defect states may be also changed when temperature changes. For further confirmation, as shown in the inset of Fig.4, we focus on the temperature dependence of the defect emission intensity. It increases with decreasing the temperature to 150 K but drops with further temperature decrease. This unusual temperature dependence is likely to be a result of the interplay between the emission efficiency associated to defect states and the thermally-activated carrier diffusion. When temperature goes down, the initial increase in emission efficiency may be understood due to the suppression of nonradiative recombination from defect states; nevertheless, the behavior at lower temperature should be associated to the thermally-activated carrier diffusion. It becomes more important when temperature increases, leading to the enhancement of defect-related recombination. Such a critical temperature is close to the onset temperature of thermally-activated carrier diffusion between weakly and strongly localized states,[28] which further confirms the validity of assignment of thermal activation. From above discussions, we can safely conclude that the thermally-activated carrier diffusion process makes contributions to the DADR, as indicated by the superlinear fluence dependence of $I_{Def}/I_{Blue}$, leading to the efficiency droop in LEDs. As mentioned above,[14] the e-e scattering can induce the DADR with a temperature-insensitive possibility. We note here that the data obtained in this study indicates the presence of the thermally-activated carrier diffusion, but doesn't exclude the contribution of the e-e scattering.

Despite this study is mainly done by monitoring the defect emission, the mechanism is also valid in samples where defect-related recombination is dominated by nonradiative processes. Different from the general SRH recombination, the possibility of DADR is not sensitive to the density of defect states but is strongly dependent on the energy barriers required to overcome. The dominant mechanism for efficiency droop has been highly controversial in spite of a great number of studies on it. Possible reasons may include the restrictions of specific experimental techniques as well as the sample diversities. For example, the PL study done here can hardly evaluate the importance of droop related to the electron leakage during current injection. For retaining the efficiency in LEDs, it is of more importance to eliminate the possible loss channels. In principle, the mechanism of efficiency droop we observed here can be reduced by manipulating the lateral carrier confinement. The carrier diffusion lengths in GaN-based QWs have been characterized to be at the order of tens of nanometers.[34] Lateral confinement with micro-fabricated nanostructures can be an effective approach to suppress it.[35,36] To avoid introducing more surface states, the top-down epitaxial growth of nano-LEDs may be a more ideal solution.[37,38]

In summary, we have observed experimental evidences in a-plane InGaN QWs for a plausible mechanism of efficiency droop as the DADR. The excitation-photon-dependent and temperature-dependent experiments separately and corporately support the thermally-activated carrier diffusion as a major process inducing the electron migration from localized states to defect states. Such a mechanism may be responsible for the residual efficiency droop in nonpolar LEDs, which can be further manipulated by nano-architecture towards droop free devices.

This work is supported by the Program of International S&T Cooperation (2011DFA01400, MOST), the National Basic Research Program of China (2012CB921801 and 2013CB932903, MOST), the National Science Foundation of China (9123310, 61108001, 11227406 and 11021403), and NRF of Korea (K2011-0017325).. The author C.Z. acknowledges financial support from New Century Excellent Talents program (NCET-09-0467), Fundamental Research Funds for the Central Universities, and the Priority Academic Program Development of Jiangsu Higher Education Institutions (PAPD). J.S.K acknowledges financial support from WCU program in SCNU.

**Figure 1**

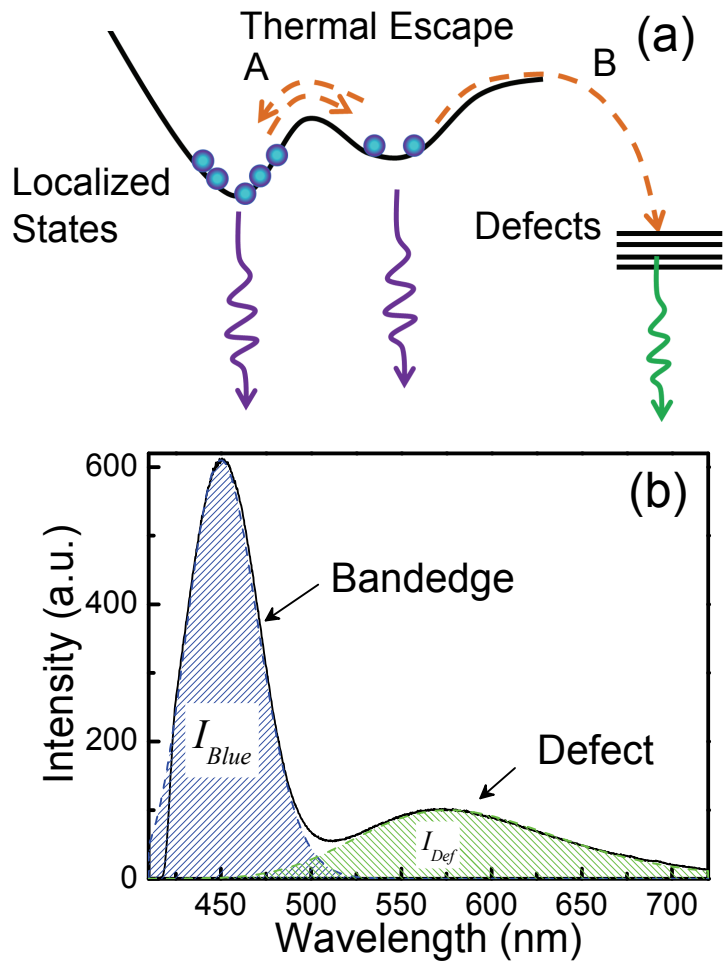

FIG. 1. (color online) (a) Schematic sketch of the DADR process. Radiative recombinations from localized states and defects contribute to the bandedge blue emission and broadband yellow emission. (b) PL spectrum from MQWs under relatively high fluence excitation. The spectrum is compared with the Gaussian profiles of a bandedge emission band (blue) and a defect emission band (green).

**Figure 2**

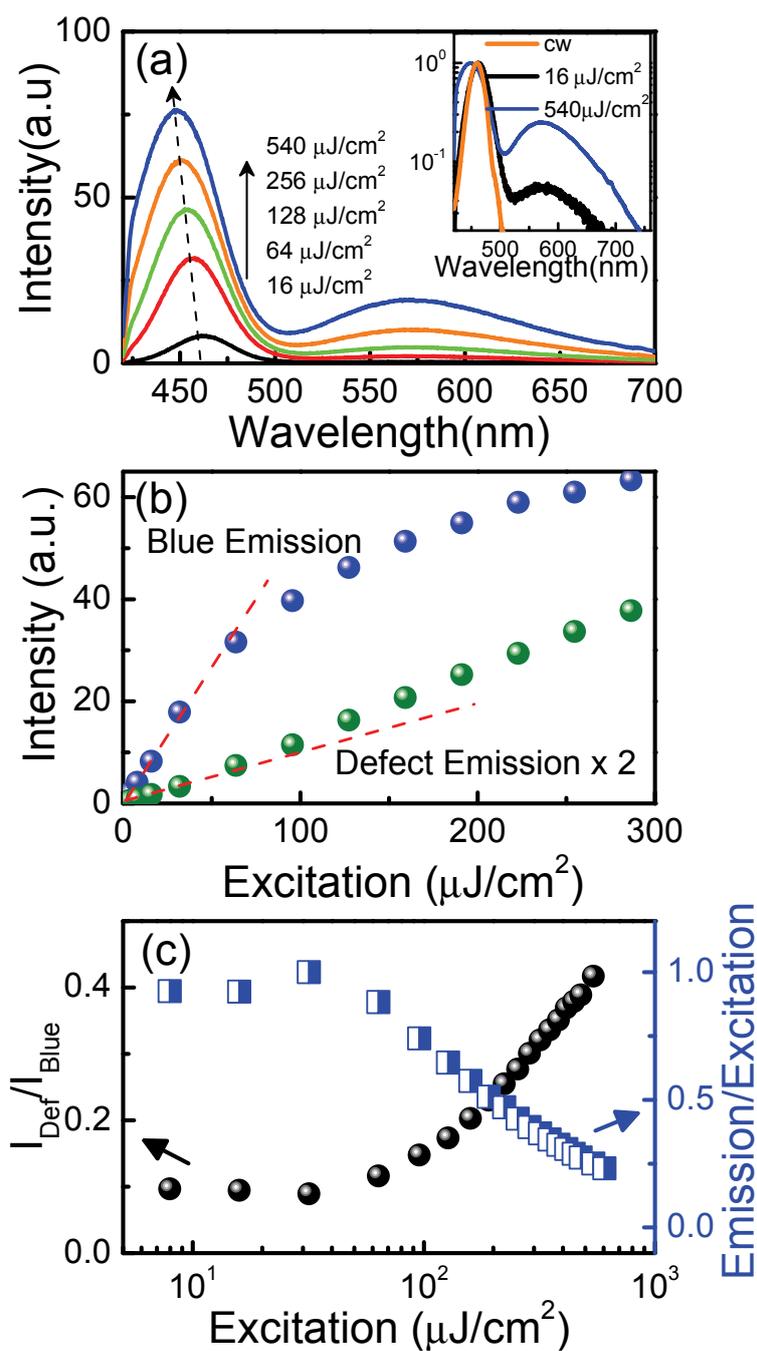

FIG.2. (color online) (a) Room-temperature emission spectra under different excitation fluences. The inset compares the normalized spectra recorded under cw, weak pulse, and intense pulse excitation, respectively. (b) The integrated intensities of blue emission (blue) and defect emission (green) are plotted versus the excitation fluence. (c) The intensity ratio ($I_{Def}/I_{Blue}$) and the normalized bandedge emission intensity per unit excitation power are plotted as a function of excitation fluence.

**Figure 3**

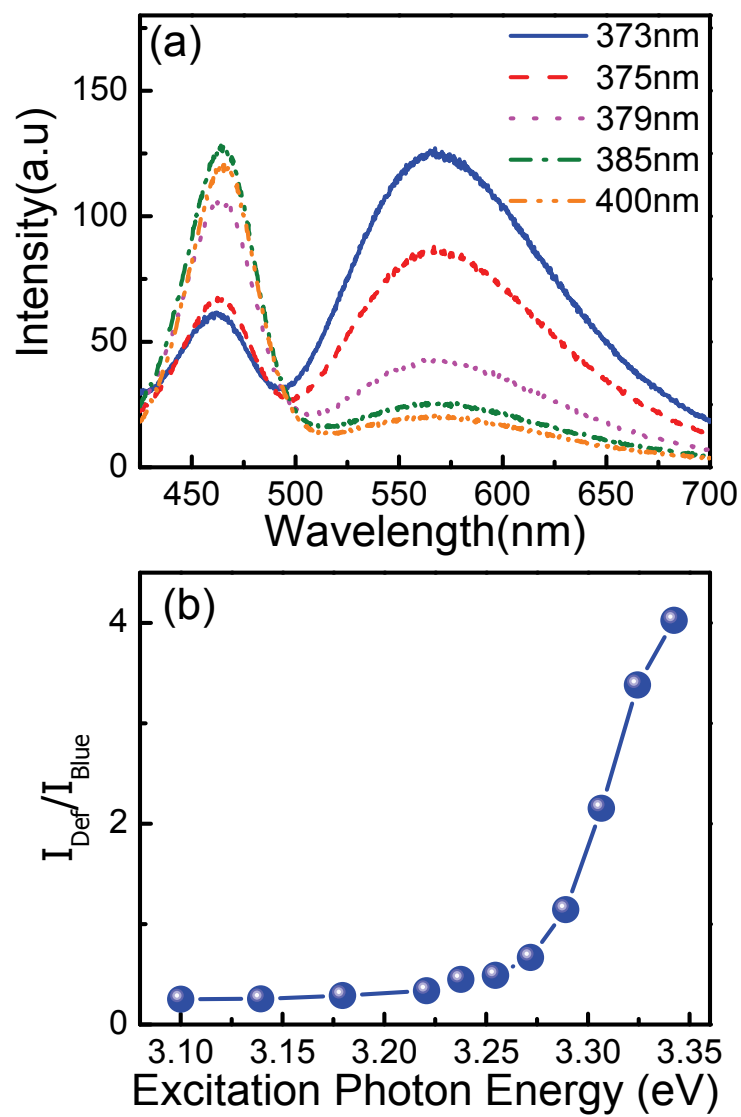

FIG. 3. (color online) (a). Room-temperature emission spectra recorded with different excitation wavelength. (b). The intensity ratio $I_{Def}/I_{Blue}$ is plotted as a function of excitation photon energy.

**Figure 4**

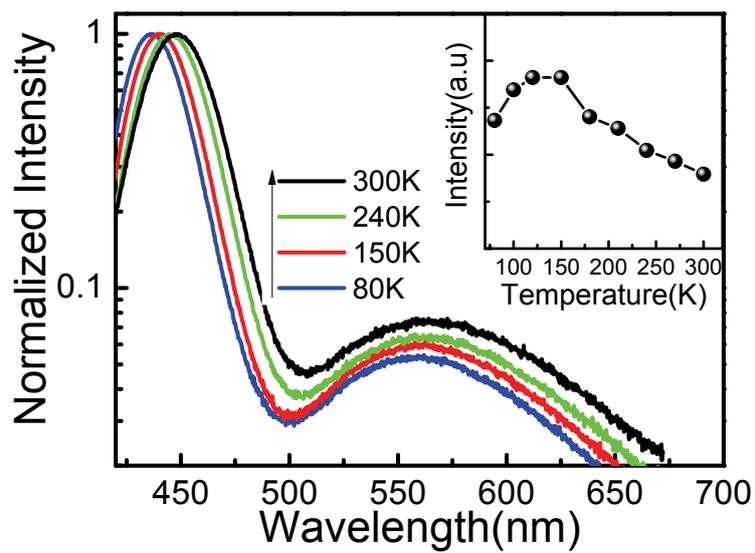

FIG. 4. (Color online) Plot of the normalized PL spectra with different temperature. The inset shows the peak intensity of defect emission versus temperature.